\documentclass[superscriptaddress,aps,journal=prl,preprint,longbibliography]{revtex4-2}
\usepackage{graphicx}
\usepackage{epstopdf}
\usepackage{booktabs}
\usepackage{tabularx}
\usepackage{ragged2e}
\usepackage{array}
\usepackage{bm,amsmath,amssymb} 
\usepackage{color, soul} 
\usepackage{dcolumn}   
\usepackage{multirow}
\usepackage{nccmath} 

\usepackage{hyperref}
\hypersetup{
    colorlinks=true,
    linkcolor=blue,
    citecolor=blue,     
    urlcolor=blue,
}

\setcitestyle{super,comma,sort&compress,open={},close={}}

\usepackage{soul}

\begin{document}
    \title{Emergent magnetic nonanalyticity in Dirac altermagnets}
   
    \author{Xizhi Fu}
    \thanks{These authors contributed equally.}
    \affiliation{Department of Physics, Hong Kong University of Science and Technology, Clear Water Bay, Hong Kong SAR, China}

    \author{Zhenqiao Huang} 
    \thanks{These authors contributed equally.}
    \affiliation{Department of Physics, Hong Kong University of Science and Technology, Clear Water Bay, Hong Kong SAR, China}

    \author{Fei Yang}
    \thanks{These authors contributed equally.}
    \affiliation{Department of Physics, Hong Kong University of Science and Technology, Clear Water Bay, Hong Kong SAR, China}
    
    \author{Mengli Hu}
    \affiliation{Department of Physics, Hong Kong University of Science and Technology, Clear Water Bay, Hong Kong SAR, China}
    
    \author{Xingkai Cheng}
    \affiliation{Department of Physics, Hong Kong University of Science and Technology, Clear Water Bay, Hong Kong SAR, China}
    
    \author{Junwei Liu}
    \email{liuj@ust.hk}
    \affiliation{Department of Physics, Hong Kong University of Science and Technology, Clear Water Bay, Hong Kong SAR, China}
    \affiliation{IAS Center for Quantum Matter, Hong Kong University of Science and Technology, Clear Water Bay, Hong Kong SAR, China}
    
	\begin{abstract}
Symmetry has long provided the organizing principle for understanding how materials respond to external perturbations. Here we show that what is permitted by symmetry need not be what actually emerges from the underlying system, and this distinction becomes particularly consequential for symmetry-related gapless Dirac fermions such as in altermagnets. A perturbation that shifts the Dirac points drives a Lifshitz reconstruction and generates an intrinsically nonanalytic effective potential, giving rise to undiscovered singular responses. In Dirac altermagnets, this nonanalyticity acquires a  magnetic manifestation through $C$-paired spin-polarized Dirac cones: breaking the pairing symmetry $C$ lifts their degeneracy and converts the associated Lifshitz reconstruction into an itinerant magnetization. As a consequence, piezomagnetism, the induction of a net magnetization by a mechanical strain, becomes nonlinear and nonanalytic, with $M\propto\mathrm{sgn}(\varepsilon)\varepsilon^2$, whereas the linear response is absent despite being allowed by symmetry. Accordingly, the leading response cannot be inferred from the conventional symmetry-based paradigm of material response built upon the cornerstones of Neumann's principle and analytic Landau expansion.  First-principles calculations identify the monolayer Fe$_2$S as a candidate material realization for such singular magnetic responses. These results suggest quantum singularities as a new degree of freedom for engineering material responses.

	\end{abstract}
	\maketitle

\newpage

Understanding how a material responds to external perturbations is a central theme of condensed-matter physics and materials science. From electric and magnetic fields to mechanical deformation and electromagnetic radiation, material responses, ranging from susceptibilities to nonlinear optical phenomena such as second-harmonic generation, are conventionally organized according to the symmetries of the underlying state \cite{Dresselhaus2008,Seemann2015}. Neumann's principle provides a fundamental basis for this description by constraining the response coefficients allowed by crystal symmetry, while Landau theory further describes material responses through analytic expansions in fields and order parameters \cite{Landau1937,ChaikinLubensky1995}. These principles have long formed a powerful framework for understanding and classifying material responses. Symmetry, however, determines what is allowed, but not necessarily what actually emerges from the underlying system, as the functional form of a macroscopic response may encode physics that cannot be inferred from symmetry considerations alone.

The discovery of material platforms hosting such qualitatively new response functions can, in principle, provide fertile ground for uncovering new concepts with far-reaching implications across science and potential industry applications. Recently, altermagnets (A$l$Ms) \cite{libor2022,bai2024,song2025,cheong2025,xie2026review} have emerged as a distinct class of magnetic materials that simultaneously possess zero magnetization and spin-split band structures \cite{wu2004,hayami2019,libor2020,linding2020,haiyang2021,mazin2021,krempasky2024,reimers2024,lee2024broken,jiang2025metallic,yang2025crsb,gu2025ferroelectric,mu2026quadruple}, and exhibit intriguing physical responses, including non-relativistic noncollinear spin currents \cite{makoto2019,haiyang2021,rafael2021,dingfu2021} and unconventional piezomagnetism (PZM) \cite{haiyang2021,aoyama2024,takahashi2025}, which distinguish them from ferromagnets and conventional antiferromagnets. These responses are rooted in strict $C$-paired spin-momentum locking (CSML) \cite{haiyang2021,mengli2025}, where $C$ denotes crystal symmetry, such as rotation and/or mirror but not translation or inversion, that interchanges opposite magnetic sublattices and thereby characterizes the A$l$M order and its deterministic switching behavior \cite{lei2024,zhou2025,yiyuan2025}. In addition to defining the symmetry constraints of A$l$M responses, CSML establishes a direct link between symmetry-breaking perturbations and the underlying electronic band structure. This connection becomes particularly consequential when A$l$M bands host distinctive electronic structures, including topological \cite{lu2025topological,radhakrishnan2026}, Dirac \cite{Bernevig2006,Zhang2009} , flat, and massless \cite{parshukov2025} bands.
Such electronic features can be imprinted onto macroscopic observables, generating qualitatively new response phenomena in contrast to conventional linear responses~\cite{haiyang2021} and providing a unique platform for exploring the interplay among spin, charge, lattice, and momentum degrees of freedom.

In this work, we propose a generic quantum-singularity-induced response in a  typical $d$-dimensional Dirac system. 
The resulting effective action acquires a nonanalytic contribution that inherits the singular scaling of the Dirac spectrum, thereby altering the analytic structure and determining the leading functional form of the macroscopic response not captured by conventional symmetry-based classifications and analytic theories. In practice, PZM in A$l$Ms provides a direct physical realization of this principle, where symmetry-breaking strain redistributes spin-polarized itinerant electrons and generates a net magnetization. Specifically, two-dimensional Dirac A$l$Ms host a previously unrecognized nonlinear and nonanalytic PZM, in which the symmetry-allowed linear response is absent and the leading magnetization follows $M\propto \mathrm{sgn}(\varepsilon)\varepsilon^2$. Based on first-principles calculations and symmetry analysis, we identify monolayer $\mathrm{Fe_2S}$ as a realistic material candidate and explicitly demonstrate this distinct PZM response. These findings establish quantum singularities as a new frontier for controlling material responses, revealing a route by which the microscopic quantum structure of matter is directly translated into macroscopic functionalities beyond symmetry principles.

\section*{Dirac model and nonanalytic response}

We consider a typical $d$-dimensional Dirac system~\cite{peskin2018introduction} coupled to a
symmetry-breaking bosonic field $\phi$. The minimal Hamiltonian can be
written as
\begin{equation}
H
=
\sum_{\zeta=\pm}
\sum_{\bf k}
\psi_{\zeta,\bf k}^{\dagger}
\bigg[
\sum_{i=1}^{d} v_i k_i \Gamma_i
+
\zeta(\phi-h)
\bigg]
\psi_{\zeta,\bf k}
+
V_{\rm B}[\phi],\nonumber
\label{eq:general_H}
\end{equation}
where $\psi_{\zeta,\bf k}$ is the Dirac fermion spinor in the $\zeta$ sector,
$\zeta=\pm$ labels the two symmetry-related Dirac sectors,
$\Gamma_i$ are mutually anticommuting Dirac matrices acting on the
internal degrees of freedom of the Dirac fermion and satisfying
$\{\Gamma_i,\Gamma_j\}=2\delta_{ij}$, and $v_i$ are the corresponding
Dirac velocities. The bosonic field $\phi$ couples oppositely to the
two sectors and therefore shifts their Dirac points in opposite
directions in energy without opening a gap. The auxiliary field $h$
couples to the same fermionic channel and is introduced as a source
field for the corresponding response, while $V_{\rm B}[\phi]$ denotes
the bare bosonic potential. 
Integrating out the gapless Dirac fermions at zero temperature generates a
nonanalytic contribution to the bosonic effective potential,
\begin{equation}
V_{\rm eff}(\phi,h)
=
V_{\rm B}(\phi)
-
\mathcal{C}_d
|\phi-h|^{d+1}
+\cdots ,\nonumber
\label{eq:Veff_general}
\end{equation}
with $\mathcal{C}_d
=
\frac{
2\mathcal{S}_{d-1}
}{
(2\pi)^d
d(d+1)
\prod_{i=1}^{d}v_i
}$, where $\mathcal{S}_{d-1}$ is the area of the $(d-1)$-dimensional unit sphere. 

Thus, the response conjugate to $h$,
$\mathcal{O}_{\phi}
=
-
\left.
\frac{\partial V_{\rm eff}}
{\partial h}
\right|_{h=0}$,
takes the universal scaling form
\begin{equation}
\mathcal{O}_{\phi}
=
-(d+1)\mathcal{C}_d
\,\phi|\phi|^{d-1}
\propto
\operatorname{sgn}(\phi)|\phi|^d.\nonumber
\label{eq:general_response}
\end{equation}
Consequently, while symmetry alone would allow the response to begin at linear order (e.g., $\mathcal{O}^{\rm c}_{\phi}=c_1\phi+c_2\phi^2+c_3\phi^3\cdots$), since $\phi$ and its conjugate
response transform in the same symmetry channel, the Dirac contribution here  scales as $\mathcal{O}_{\phi}\propto
\operatorname{sgn}(\phi)|\phi|^d$.
In other words, Dirac fermions impose a characteristic
dimension-dependent scaling on the conjugate response, generating
nonanalytic terms that escape a conventional analytic expansion. In particular, this gives $\mathcal{O}_{\phi}\propto\phi|\phi|$ in 2D systems, so that the
symmetry-allowed linear term is absent and the leading Dirac response
is intrinsically nonanalytic. 

The absence at an order allowed by Neumann's principle cannot be anticipated from symmetry, and the nonanalytic form cannot be captured by a conventional analytic Landau expansion.  The underlying origin lies in the dimensionality and singular structure of the electronic spectrum of a gapless Dirac fermion, which constitutes a Lifshitz point and
an infinitesimal energy shift of the Dirac cone changes the Fermi
surface from a point to a finite Fermi pocket. 
Such singularity is directly transferred into the bosonic effective theory, thereby imprinting the characteristic dimensional scaling of the Dirac spectrum onto the effective potential. This changes not merely the coefficients of the response but its analytic structure. Such a mechanism may provide a route by which microscopic quantum singularities can reshape the relation between symmetry and macroscopic response, allowing the functional form of the response itself to encode information not captured by conventional symmetry-based and analytic descriptions.

\begin{figure}[!htbp]
\includegraphics[width=0.8\linewidth]{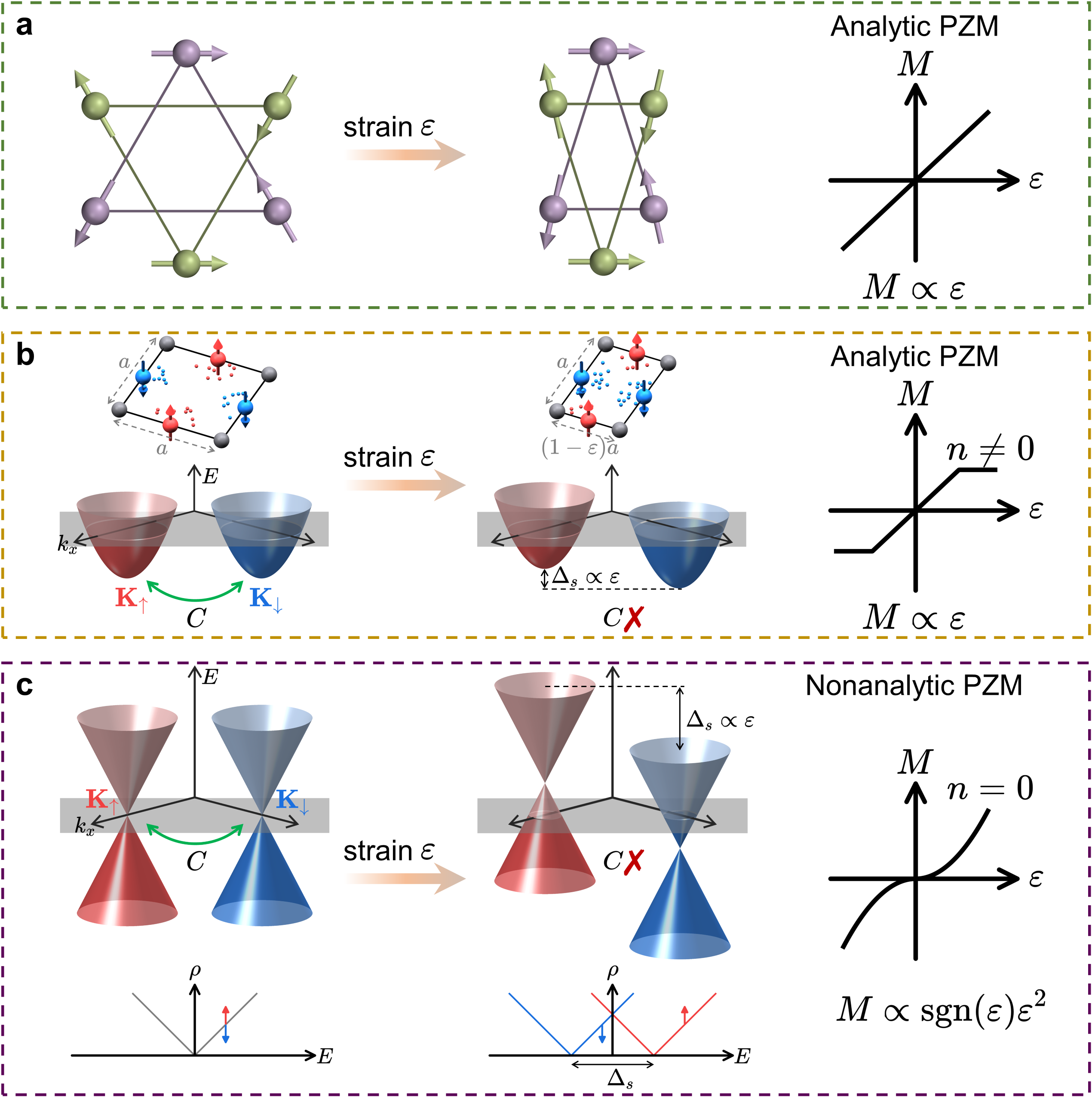}
\caption{\label{fig_1}
\textbf{Three mechanisms for PZM.} 
\textbf{a,} Conventional PZM arising from strain-induced reorientation of local magnetic moments \cite{ikhlas2022}. 
\textbf{b,c,} PZM arising from strain-induced redistribution of itinerant electrons in a metallic A$l$M \cite{haiyang2021} and a 2D Dirac A$l$M, respectively. 
The induced magnetization $M$ scales as $M\propto\varepsilon$ in \textbf{a} and \textbf{b} and $M\propto \mathrm{sgn}(\varepsilon)\varepsilon^2$ in \textbf{c}, respectively.
Note that conventional analytic PZM can generally be expanded as $M=\sum_{m=1}^{\infty}c_m\varepsilon^m$, whereas the scaling $M\propto \mathrm{sgn}(\varepsilon)\varepsilon^2$ in \textbf{c} is nonanalytic at $\varepsilon=0$.
In \textbf{a} and \textbf{b}, the spheres and arrows denote atoms and local magnetic moments, where red/blue dots in \textbf{b} represent spin-up/spin-down itinerant electrons.
In \textbf{b} and \textbf{c}, the red and blue surfaces denote opposite-spin energy bands at $\mathbf{K}_\uparrow$ and $\mathbf{K}_\downarrow$, related by the crystal symmetry $C$; strain breaking $C$ induces a relative energy shift $\Delta_s\propto\varepsilon$; $n$ denotes the carrier density. 
The lower density of states (DOS) sketches in \textbf{c} illustrate that the linear nonanalytic Dirac DOS leads to the nonlinear and nonanalytic PZM response.
}
\end{figure}

\section*{Nonanalytic piezomagnetism}

A natural realization of the symmetry-breaking bosonic field $\phi$ is a mechanical strain $\varepsilon$, through which the Dirac nonanalyticity acquires a direct magnetic manifestation as PZM in 2D Dirac A$l$Ms. Specifically, as illustrated in Fig.~\ref{fig_1}c, we consider a pair of Dirac cones  $D_\sigma\ (\sigma = \uparrow,\downarrow)$ with opposite spins centered at $\mathbf{K}_\uparrow$ and $\mathbf{K}_\downarrow$, related by a crystal symmetry $C$ as $\mathbf{K}_\uparrow=C\mathbf{K}_\downarrow$. As long as $C$ is preserved, the two Dirac cones remain degenerate and equally occupied, leading to full spin compensation and hence zero net magnetization. 

For a fixed strain channel, $\varepsilon_{ij}=e_{ij}\varepsilon$, a strain component $\varepsilon$ that breaks $C$ induces a relative energy shift $\Delta_s\propto\varepsilon$ between the two opposite-spin Dirac cones without opening a gap \cite{bir1974,haiyang2021}.
This strain-induced energy shift plays the role of the symmetry-breaking field $\phi$. The resulting occupation imbalance between the two spin-polarized Dirac sectors generates a net itinerant magnetization, $M\propto n_\uparrow-n_\downarrow$.  Then, as expected from the generic behavior of 2D Dirac A$l$Ms discussed above, the strain-induced occupation imbalance between the two opposite-spin Dirac cones yields $M\propto \mathrm{sgn}(\varepsilon)\varepsilon^2$. 
Consequently, although the relative energy shift between the two Dirac cones is linear in strain, the resulting magnetization does not exhibit the conventional analytic linear PZM response allowed by symmetry, and the itinerant-electron mechanism gives rise to a nonlinear and nonanalytic PZM response.

Strain can also alter the Dirac group velocity or even gap out the Dirac cones. To capture these effects on equal footing, we introduce a general 2D  four-component low-energy Hamiltonian describing a pair of symmetry-related, spin-polarized Dirac cones,
\begin{equation}
H_0
=
\sum_{\mathbf k}
\Psi_{\mathbf k}^{\dagger}
\Big[
\sum_{i=x,y}\sum_{a=0,x,y,z}
k_i\,\hat{\mathcal V}_{i,a}\tau_a
\Big]
\Psi_{\mathbf k},
\nonumber
\end{equation}
where  $\hat{\mathcal V}_{i,a}
=
v_{i,a}^{(0)}\sigma_0
+
v_{i,a}^{(s)}\sigma_z$ acts in the spin sector, with the dispersions of the two opposite-spin Dirac cones constrained by the crystal symmetry $C$ of A$l$Ms that maps one cone onto the other. 
Here the superscripts $(0)$ and $(s)$ denote the spin-independent and -dependent components, respectively.
The four-component spinor  
$\Psi_{\mathbf k}^{\dagger}
=(
\psi_{\uparrow 1,\mathbf k}^{\dagger},
\psi_{\uparrow 2,\mathbf k}^{\dagger},
\psi_{\downarrow 1,\mathbf k}^{\dagger},
\psi_{\downarrow 2,\mathbf k}^{\dagger})$, where $1,2$ label the two low-energy orbital states forming each Dirac cone. Here $\tau_a$ act in the two-orbital Dirac subspace, with the $a=0$ terms describing the possible tilt of the Dirac cones and the $a=x,y,z$ terms determining the Dirac pseudospin texture, whereas $\sigma_i$ act in the physical-spin space. To first order in strain, 
 the momentum-dependent effect is incorporated through $\mathbf{k} \to \mathbf{k}^{\varepsilon} \equiv (\mathbf{1}-2\bm{\varepsilon})\mathbf{k}$, whereas the momentum-independent deformation potential takes the general form
\begin{align}
H_{\varepsilon}
=
\sum_{\bf k}\sum_{i,j=x,y}
\varepsilon_{ij}\,
\Psi_{\bf k}^{\dagger}
\Big[
\Lambda_{ij}^{(0)}\sigma_0\tau_0
+
\Lambda_{ij}^{(s)}\sigma_z\tau_0+
\sum_{a=x,y,z}
\left(
\Gamma_{ij,a}^{(0)}\sigma_0
+
\Gamma_{ij,a}^{(s)}\sigma_z
\right)\tau_a
\Big]
\Psi_{\bf k}.\nonumber
\label{eq:Hstrain_general}
\end{align} 
Within the spin-group description \cite{liu2022spingroup,chen2024spin,zeng2024}, two classes of symmetries are essential: symmetries $G_\sigma$ in the little groups of $\mathbf{K}_\sigma$ and symmetries $C$ interchanging $\mathbf{K}_\uparrow$ and $\mathbf{K}_\downarrow$. 
The symmetries $G_\sigma$ protect the gapless Dirac crossings by forbidding mass terms, and the symmetries $C$ map $D_\uparrow$ and $D_\downarrow$ to each other. Breaking $G_\sigma$ while preserving $C$ gaps the two cones without generating a net magnetization, whereas breaking all $C$ symmetries while preserving $G_\sigma$ shifts the two cones relative to each other and produces spin imbalance. A generic strain may induce both effects simultaneously.

\begin{figure}[!tbp]
\includegraphics[width=0.6\linewidth]{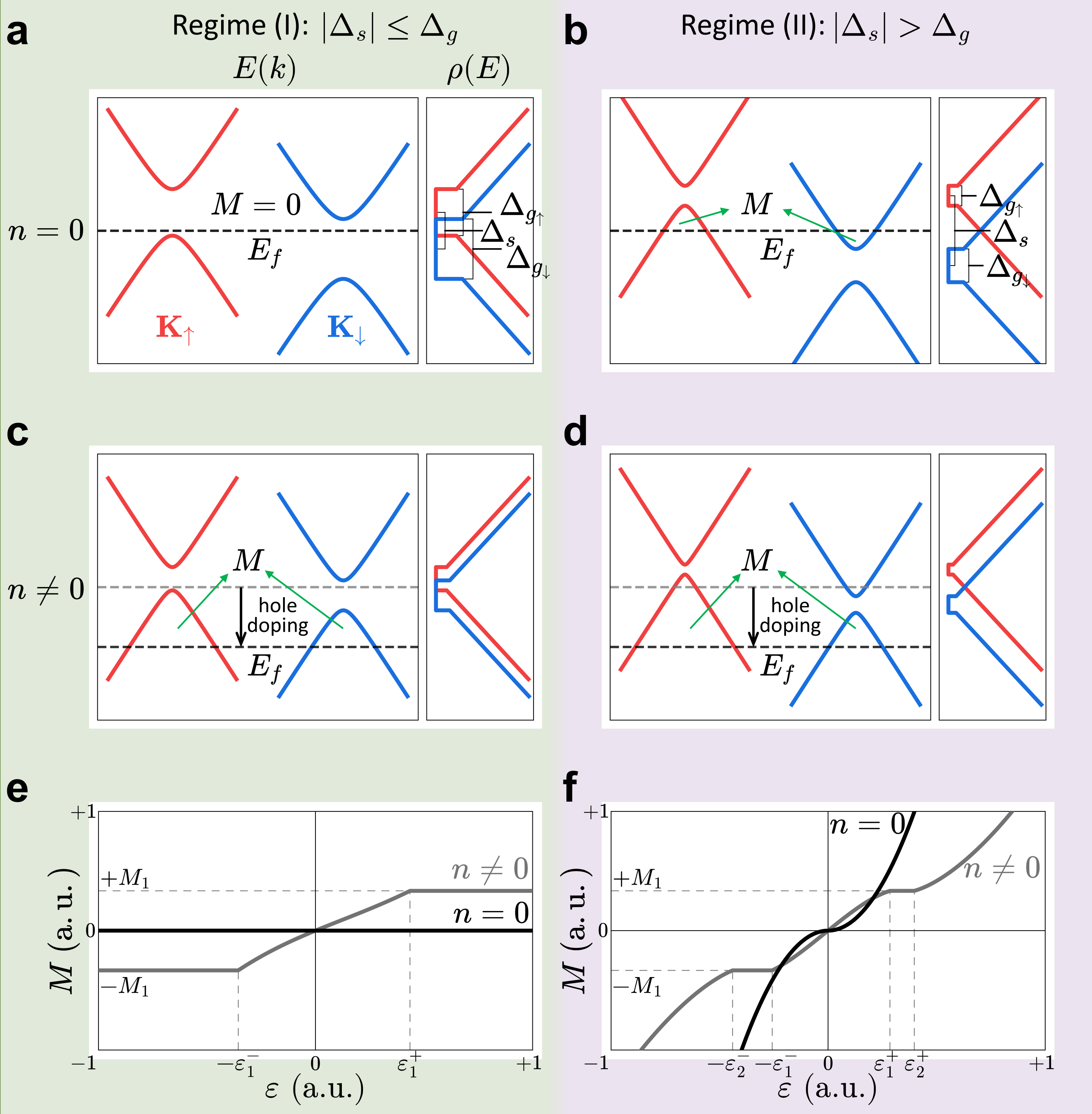}
\caption{\label{fig_2} 
\textbf{$k \cdot p$-model schematics for strained 2D Dirac A$l$Ms.}
\textbf{a--d,} The four panels illustrate regimes defined by the relative energy shift $\Delta_s$, band gap $\Delta_g$, and doping-induced carrier density $n$:
\textbf{a,} $|\Delta_s|\leq\Delta_g$ and $n=0$,
\textbf{b,} $|\Delta_s|>\Delta_g$ and $n=0$,
\textbf{c,} $|\Delta_s|\leq\Delta_g$ and $n\neq0$, and
\textbf{d,} $|\Delta_s|>\Delta_g$ and $n\neq0$.
In each panel, the band structure is at left and the corresponding DOS is at right.
Red and blue represent the spin-up states at $\mathbf{K}_\uparrow$ and spin-down states at $\mathbf{K}_\downarrow$, respectively.
Black dashed lines denote the Fermi level $E_f$,
gray dashed lines mark the reference energy of the unstrained Dirac points,
green arrows indicate the contribution to net magnetization $M$,
and black vertical arrows indicate hole doping.
\textbf{e,f,} The dependence of net magnetization $M$ on signed strain amplitude $\varepsilon$ for $|\Delta_s|\leq\Delta_g$ and $|\Delta_s|>\Delta_g$, respectively.
}
\end{figure}

At zero temperature, the strain-modified Dirac cones retain the characteristic linear DOS of Dirac fermions (see Supplementary Note 2 for detailed derivations): 
\begin{equation}
\label{eq_dos}
\begin{aligned}
\rho_\uparrow(E)=
\begin{cases}
\alpha\left|E\!-\!\frac{1}{2}\Delta_s \right|, & |E \!-\!\frac{1}{2}\Delta_s|\geq  \frac{1}{2}\Delta_{g_\uparrow}\\[2pt]
0, & |E\!-\!\frac{1}{2}\Delta_s|<  \frac{1}{2}\Delta_{g_\uparrow}
\end{cases}\quad
\rho_\downarrow(E)=
\begin{cases}
\alpha\left|E\!+\!\frac{1}{2}\Delta_s \right|, & |E \!+\!\frac{1}{2}\Delta_s|\geq  \frac{1}{2}\Delta_{g_\downarrow}\\[2pt]
0, & |E\!+\!\frac{1}{2}\Delta_s|<  \frac{1}{2}\Delta_{g_\downarrow} \nonumber
\end{cases}
\end{aligned},
\end{equation}
where $\alpha>0$ is a system-dependent constant, 
$\Delta_s = s(\theta)\varepsilon$ denotes the relative energy shift between the two Dirac cones induced by a strain of signed amplitude $\varepsilon$ and polar angle $\theta$, and $\Delta_{g_\sigma}=g_\sigma(\theta)|\varepsilon|$ denotes the corresponding strain-induced gaps of the Dirac cones $D_\sigma$.

Then, as summarized in Fig.~\ref{fig_2}, the resulting PZM behavior is controlled by the competition between the strain-induced relative energy shift $\Delta_s$ and the average strain-induced band gap $\Delta_g=(\Delta_{g_\uparrow}+\Delta_{g_\downarrow})/2$, which naturally classifies strained 2D Dirac A$l$Ms into two regimes: (I) $|\Delta_s|\leq\Delta_g$ and (II) $|\Delta_s|>\Delta_g$ (including gapless strained systems with $\Delta_g=0$).  As shown in Fig.~\ref{fig_2}a, in regime (I) at $n=0$, the Fermi level lies within the band gaps, preventing strain from inducing a net magnetization, as illustrated by the black $n=0$ curve in Fig.~\ref{fig_2}e. This corresponds to a compensated ferrimagnet, where the net magnetization vanishes due to equal occupations of opposite-spin electronic states. By contrast, in regime (II) at $n=0$, the Fermi level crosses the bands of both Dirac cones, resulting in an occupation imbalance and a finite net magnetization, as shown in Fig.~\ref{fig_2}b. Imposing conservation of electron number, the net static magnetization (black $n=0$ curve in Fig.~\ref{fig_2}f) is obtained as (see Supplementary Note 2 for detailed derivations):
\begin{equation}
M(\varepsilon)=\beta(\alpha,g_{\sigma},s)\,\mathrm{sgn}(\varepsilon)\varepsilon^2,
\nonumber
\end{equation}
exhibiting nonlinear and nonanalytic strain dependence.

As the strain amplitude changes from a positive value through zero to a negative value, the Fermi contours of the Dirac cones evolve from loop-like to point-like and then back to loop-like, as shown in Supplementary Fig.~1. At the Lifshitz point ($\varepsilon=0$), the Fermi-surface topology changes between a finite Fermi pocket and a Dirac point, where the nonanalytic Dirac DOS generates a nonanalytic contribution to the thermodynamic potential that unambiguously characterizes the Lifshitz transition (see Supplementary Note 3 for details). The nonlinear and nonanalytic PZM therefore provides a direct magnetic manifestation of this strain-induced Lifshitz transition.

For the finite-doping case [Figs.~\ref{fig_2}c and d], the resulting net magnetization is shown by the gray $n\neq0$ curves in Figs.~\ref{fig_2}e and f (see Supplementary Eq.~(S75) for the full expression). Finite doping introduces a characteristic crossover scale in the PZM response. At small strain, the response is dominated by the redistribution of doped carriers near the Fermi level, resulting in a conventional linear PZM. With increasing strain, the doped carriers are progressively polarized into one Dirac cone and eventually become fully polarized, producing a magnetization plateau whose magnitude is proportional to the carrier density. In regime (II), further increasing $|\varepsilon|$ drives the response beyond the regime dominated by doped carriers. For sufficiently large $|\varepsilon|$, the magnetization follows 
$M(\varepsilon)
=
\mathrm{sgn}(\varepsilon)\left(\beta\varepsilon^2-\eta\varepsilon^{-2}\right)+M_2$, where $\eta$ and $M_2$ are constants. In this regime, the nonlinear and nonanalytic contribution $\beta\,\mathrm{sgn}(\varepsilon)\varepsilon^2$ gradually dominates the strain dependence. Finite carrier doping and finite temperature introduce additional energy scales, $E_f$ and $k_BT$, respectively, that compete with the strain-induced Dirac energy shift. Both effects generate a crossover in the PZM response from a conventional carrier- or thermally-dominated regime at small strain to the intrinsic nonlinear and nonanalytic Dirac response when the strain-induced splitting exceeds these additional energy scales. Neither finite doping nor finite temperature, however, eliminates the intrinsic nonanalytic contribution originating from the singular Dirac electronic structure (see Supplementary Note 2 for details).

\section*{Material realization}

\begin{figure}[!tbp]
\includegraphics[width=0.7\linewidth]{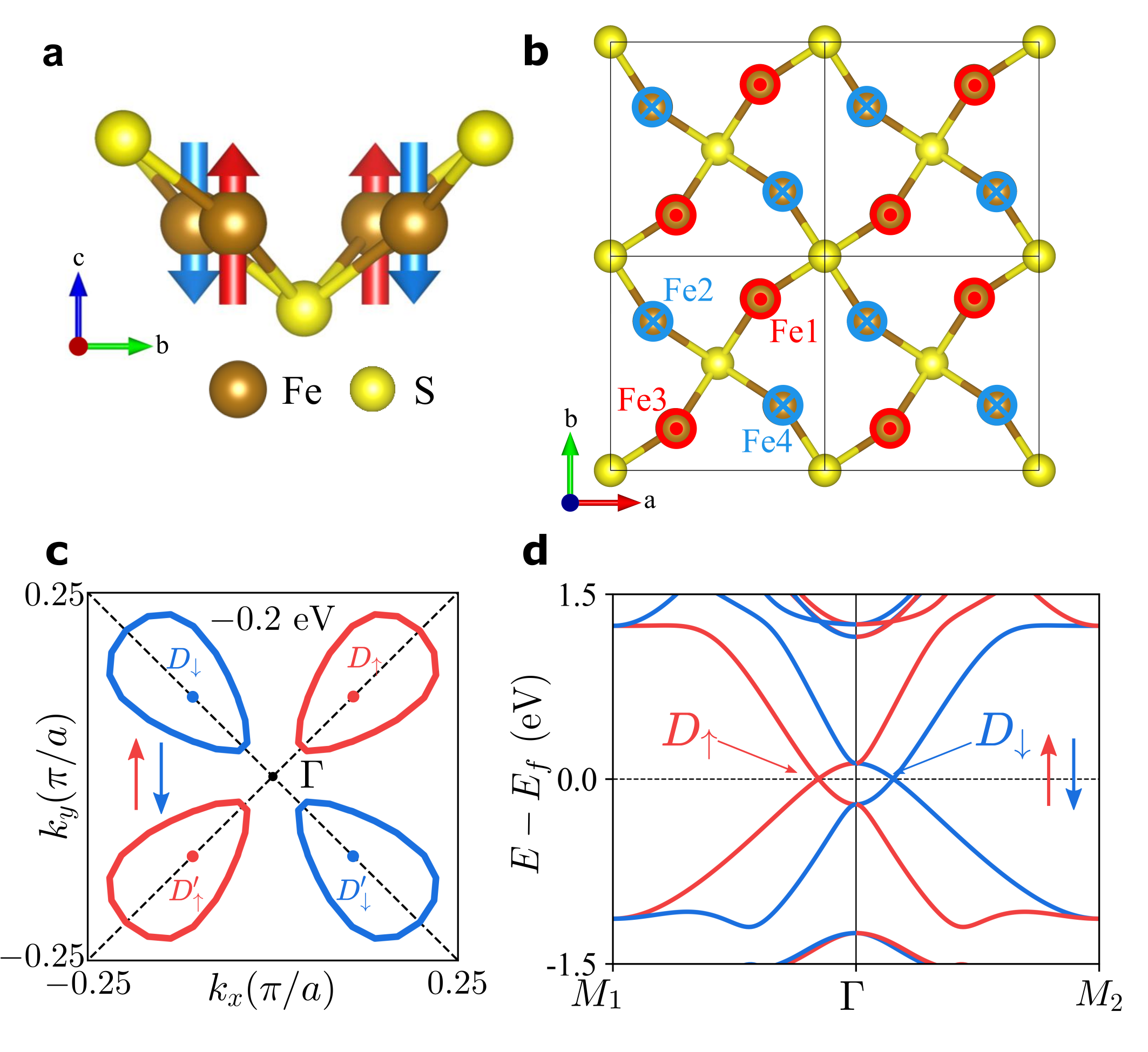}
\caption{\label{fig_3}
        \textbf{Crystal and electronic structures of monolayer Fe$_2$S.}
		\textbf{a,b,} Side and top views of monolayer Fe$_2$S, respectively, where the red and blue arrows on the Fe atoms represent opposite local magnetic moments.
		\textbf{c,} Energy contour at $0.2$ eV below the valence band maximum, where spin polarizations are denoted by red (spin-up) and blue (spin-down).
		\textbf{d,} Band structure of monolayer Fe$_2$S. In \textbf{c} and \textbf{d}, relativistic spin-orbit coupling (SOC) is neglected, and the two Dirac cones are labeled $D_\uparrow$ and $D_\downarrow$.
        }
\end{figure}

For material realization, we identify
monolayer Fe$_2$S as a realistic candidate for the nonanalytic response in $C$-paired Dirac A$l$Ms, based on first-principles calculations within density-functional theory
(DFT) \cite{hohenberg1964dft,kohn1965dft}. As shown in
Figs.~\ref{fig_3}a and b,  monolayer Fe$_2$S crystallizes in a 2D square lattice with four Fe atoms
and two S atoms in one unit cell, distributed over three atomic layers.
We choose the Cartesian axes $x$, $y$, and $z$ along the crystallographic
$[100]$, $[010]$, and $[001]$ directions, respectively. The four Fe atoms
lie in the same atomic layer and can be classified into two sublattices,
$A$ (Fe1 and Fe3) and $B$ (Fe2 and Fe4), carrying opposite local magnetic
moments along $[001]$ and forming a N\'{e}el order. The two
sublattices are connected by the symmetries $C_{4,[001]}\mathcal{T}$,
a four-fold rotation about $[001]$ followed by time reversal
$\mathcal{T}$, and $C_{2,[100]}\tau$ / $C_{2,[010]}\tau$, two-fold screw
rotations about $[100]$ / $[010]$ accompanied by the translation
$\tau=[a/2,a/2,0]$, where $a$ is the lattice constant.

\begin{figure}[!htbp]
\includegraphics[width=0.65\linewidth]{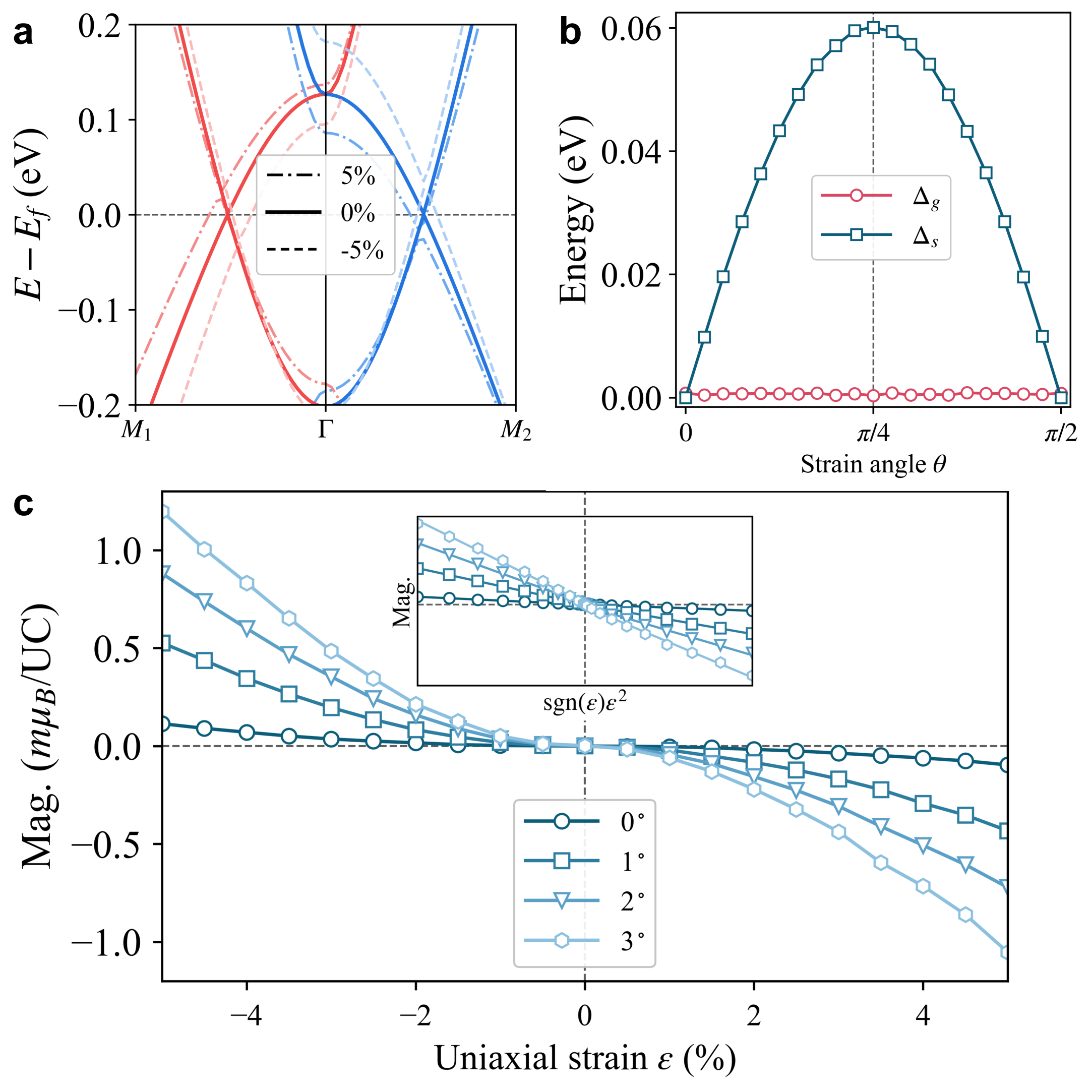}
\caption{\label{fig_4}
        \textbf{Strain-induced relative energy shift and nonlinear and nonanalytic PZM in monolayer Fe$_2$S.}
		\textbf{a,} Band structure evolution under different uniaxial strains along the [110] direction (-5\%, 0\%, and 5\%), where the two Dirac cones shift oppositely in energy, with their ordering reversed between compressive and tensile strain.
		\textbf{b,} $\Delta_g$ and $\Delta_s$ as functions of the strain direction angle ($\theta$) for 5\% strain.
		\textbf{c,} For different strain directions, the net magnetization varies nonlinearly and nonanalytically with the strain amplitude, where the inset shows fitting lines for the net magnetization versus $\mathrm{sgn}(\varepsilon)\varepsilon^2$.
        }
\end{figure}

As illustrated in Figs.~\ref{fig_3}c and d, two sets of opposite-spin energy bands in momentum space are related by the crystal symmetries $C_{4,[001]}$ and $C_{2,[100]}$ / $C_{2,[010]}$, giving rise to CSML and forming two type-I Dirac cones \cite{chang2017} $D_{\uparrow,\downarrow}$ along the $\Gamma-M_{1,2}$ paths. On each high-symmetry line, the symmetries of the corresponding little group jointly constrain the hybridization between the two low-energy bands and forbid a mass term in the Hamiltonian, thereby protecting the gapless Dirac crossing (see Supplementary Notes 4 and 5 for the symmetry analysis and first-principles verification, respectively). Moreover, relativistic spin-orbit coupling (SOC) opens only a small gap of several meV near the Dirac points, while leaving the overall linear dispersion essentially unchanged.

We then apply a compressive/tensile uniaxial strain along $[110]$, which breaks all the $C$ symmetries connecting the two opposite-spin Dirac cones, $C_{4,[001]}$ and $C_{2,[100]}$ / $C_{2,[010]}$, as illustrated in Fig.~\ref{fig_4}a. As a result, the two Dirac cones split in energy, with $D_\uparrow$ lying lower/higher than $D_\downarrow$ under compressive/tensile strain. Meanwhile, the little-group symmetries protecting the Dirac crossings remain preserved, such that the cones remain gapless. Even for strain applied along other directions, only a weak gap opening is induced, as demonstrated in Supplementary Note 5. Consequently, the criterion $|\Delta_s|>\Delta_g$ is always satisfied, as shown in Fig.~\ref{fig_4}b, placing monolayer Fe$_2$S in regime (II) for any strain considered here.

Performing first-principles calculations using dense
$\bf{k}$-grids around the Dirac points, the net magnetization is obtained by integrating
the difference between the occupied states of opposite spins in the
absence of SOC. As shown in Fig.~\ref{fig_4}c, the resulting
magnetization exhibits the predicted nonlinear and nonanalytic strain
dependence,
$M\propto\mathrm{sgn}(\varepsilon)\varepsilon^2$, in sharp contrast to
the conventional linear PZM. To the best of our knowledge, monolayer
Fe$_2$S is the first material candidate predicted to realize such
nonlinear and nonanalytic PZM.
The same mechanism may also be realized in other 2D Dirac A$l$M candidates proposed previously \cite{xu2026}.

\section*{Discussion and outlook}

The conventional theory of material responses is built upon two fundamental assumptions: symmetry determines the allowed response channels, and the leading dependence on external perturbations can be described by an analytic expansion. For example, in monolayer Fe$_2$S studied here, the crystal symmetries $C_{4,[001]}$, $C_{2,[100]}$, and $C_{2,[010]}$, which interchange the two magnetic sublattices, allow a linear PZM coupling. Accordingly, conventional Landau theory yields the lowest-order free energy $F(N,M,\bm{\varepsilon})=
\alpha_N N^2+\beta_N N^4+\alpha_M M^2
+\lambda_{xx}(\varepsilon_{xx}+\varepsilon_{yy})(N^2-N_0^2)
+\gamma\varepsilon_{xy}NM$, and generically yields a linear PZM response, $M\propto\varepsilon_{xy}$. However, our results reveal a fundamental limitation of this framework in quantum materials with singular electronic structures. In Fe$_2$S, this symmetry-allowed analytic contribution is absent in the microscopic theory without requiring any additional symmetry constraints. The gapless Dirac fermions generate a nonanalytic contribution associated with the strain-induced Lifshitz transition, transferring the singular scaling of the electronic spectrum into the macroscopic effective action and leading to 
$M\propto\mathrm{sgn}(\varepsilon)\varepsilon^2$, a breakdown of the conventional analytic expansion caused by the singular quantum spectrum.
The nonlinear and nonanalytic PZM discovered here is also fundamentally distinguished from the conventional nonlinear but analytic PZM proposed for $g$- and $i$-wave A$l$Ms \cite{mcclarty2024,ogawa2025,khodas2026}, which arises from symmetry-enforced higher-order strain couplings and remains within the description of the conventional Landau theory.

These results suggest quantum singularities as an additional organizing principle for material responses, complementary to symmetry. While symmetry specifies which responses are allowed, the microscopic quantum structure determines whether these responses emerge and the functional form of their leading contributions. A$l$Ms occupy a particularly intriguing position within this principle, because their defining symmetry $C$ relates opposite-spin quantum states at distinct momenta while enforcing exact magnetic compensation. Their magnetic symmetry gives rise to an intrinsic connection between symmetry and momentum-space quantum structure, making A$l$Ms a natural setting in which the interplay between these two organizing principles can become manifest in macroscopic responses.

The same principle may apply more generally to quantum systems with gapless electronic singularities. Whenever an external perturbation couples to such a singular electronic structure in an appropriate symmetry channel, the corresponding macroscopic response can inherit the nonanalytic scaling of the underlying quantum states. For example, a magnetic field $B$ oriented along the appropriate spin-polarization direction can couple through Zeeman interaction and induce a relative energy shift between opposite-spin Dirac cones. In a $d$-dimensional Dirac system, the singular DOS then gives rise to a singular magnetic response,  
\begin{equation}
M\propto \mathrm{sgn}(B)|B|^d\xrightarrow[]{\text{2D Dirac AlMs}}M\propto \mathrm{sgn}(B)B^2,\nonumber
\end{equation}
in sharp contrast to the conventional linear magnetic response. Accordingly, the magnetic susceptibility itself acquires a nontrivial field dependence,
$\chi(B)\propto |B|^{d-1}$,
rather than approaching a conventional constant value. Notably, the same mechanism can also operate in other response channels, including nonlinear transport, Hall effects \cite{lei2025}, and optical susceptibilities, whenever external perturbations couple to electronic singularities such as Dirac points, Van Hove singularities, flat bands, or other nonanalytic quantum spectra. In these systems, the dimensionality and singular structure of the microscopic quantum states become directly encoded in the scaling form of macroscopic observables. Identifying and controlling such quantum-singularity-driven responses may provide a new route toward quantum material design, where the quantum structure of matter itself becomes a design parameter for engineering emergent macroscopic functionalities not specified by symmetry alone.

	 

\section*{Methods}
\subsection*{First-principles calculations}
The first-principles calculations were performed within the framework of DFT as implemented in the Vienna Ab initio Simulation Package (VASP) \cite{kresse1996vasp} using the projector-augmented-wave (PAW) method \cite{blochl1994}.
The Perdew-Burke-Ernzerhof functional \cite{perdew1996} within the generalized gradient approximation (GGA) was employed to describe the exchange-correlation interaction.
The GGA+U method was employed to treat the Fe $3d$ orbitals \cite{liechtenstein1995}, where the Hubbard U was set to 4.9 eV.
The PAW potentials were used with the plane-wave energy cutoff set at 600 eV.
The Brillouin zone was sampled with a grid of $11\times11\times1$ by employing the $\Gamma$-centered Monkhorst-Pack scheme, and the convergence criteria for the electronic iterations were set to $10^{-6}$ eV.
The crystal structure was fully relaxed until the forces on each atom were smaller than $0.001~\text{eV/\AA}$.

\subsection*{General 2D strained Dirac model}
We provide a concise derivation of the DOS based on the general strained Dirac Hamiltonian introduced in the main text. In the absence of SOC, the two spin sectors are decoupled and can be treated separately. For the spin-up Dirac cone $D_\uparrow$, setting $\sigma_z=+1$ gives
\begin{equation}
c_{ia}
=
v_{i,a}^{(0)}+v_{i,a}^{(s)},
\qquad
i=x,y,\quad a=0,x,y,z,
\nonumber
\end{equation}
for the coefficients of the momentum-dependent terms. Similarly, the strain-induced deformation potentials are
\begin{equation}
\eta_{ij,0}
=
\Lambda_{ij}^{(0)}+\Lambda_{ij}^{(s)},
\qquad
\eta_{ij,a}
=
\Gamma_{ij,a}^{(0)}+\Gamma_{ij,a}^{(s)}
\quad (a=x,y,z).
\nonumber
\end{equation}
Defining
\begin{equation}
d_a
=
\sum_{i,j=x,y}
\varepsilon_{ij}\eta_{ij,a},
\qquad
\mathbf{k}^{\varepsilon}
=
(\mathbf{1}-2\bm{\varepsilon})\mathbf{k},
\nonumber
\end{equation}
the eigenvalues of the spin-up sector are
\begin{equation}
E_{\pm}
=
k_x^{\varepsilon}c_{x0}
+k_y^{\varepsilon}c_{y0}
+d_0
\pm
\sqrt{
\sum_{a=x,y,z}
\left(
k_x^{\varepsilon}c_{xa}
+k_y^{\varepsilon}c_{ya}
+d_a
\right)^2
}.
\nonumber
\end{equation}
Here, the terms proportional to $\tau_0$ describe the tilt and energy shift of the Dirac cone, whereas the terms proportional to $\tau_{x,y,z}$ determine its Dirac pseudospin structure. Under strain, the components of $\mathbf d=(d_x,d_y,d_z)$ lying within the plane spanned by the momentum-dependent Dirac vectors can be absorbed into a shift of the Dirac point, while the remaining component gives rise to a mass gap. Performing a linear transformation of momentum coordinates can  bring the momentum-dependent part into its canonical form and shift the momentum origin to the strain-modified Dirac point. The resulting dispersion can be written as
\begin{equation}
E_{\pm}
=
v_{x0}k_x^{\varepsilon}
+
v_{y0}k_y^{\varepsilon}
+
E_0
\pm
\sqrt{
(v_xk_x^{\varepsilon})^2
+
(v_yk_y^{\varepsilon})^2
+
m^2
},
\nonumber
\end{equation}
where $v_{x0}$ and $v_{y0}$ characterize the tilt of the Dirac cone, $v_x$ and $v_y$ are the corresponding Dirac velocities, $E_0$ is the strain-induced shift of the Dirac-point energy, and $m$ is the strain-induced Dirac mass, while $\frac{v_{x0}^2}{v_x^2}
+
\frac{v_{y0}^2}{v_y^2}
<1$ to ensure the type-I Dirac cone \cite{chang2017}.

Then, we make a linear transformation
 \begin{equation}
	\left( \begin{array} { l } { k_x^{\varepsilon} } \\ { k_y^{\varepsilon}} \end{array} \right) = \left( \begin{array} { c c } { \frac { 1 } { v _ { x } } \frac { 1 } { \sqrt { 1 - \lambda _ { x } ^ { 2 } } } } & { \frac { 1 } { v _ { x } } \frac { 1 } { \sqrt { 1 - \lambda _ { x } ^ { 2 } } } } \\ { \frac { 1 } { v _ { y } } \frac { 1 } { \sqrt { 1 - \lambda _ { y } ^ { 2 } } } } & { \frac { 1 } { v _ { y } } \frac { - 1 } { \sqrt { 1 - \lambda _ { y } ^ { 2 } } } } \end{array} \right) \left( \begin{array} { l } { k_a^{\varepsilon} } \\ { k_b^{\varepsilon}} \end{array} \right),\label{transA}\nonumber
    \end{equation}
    where $\lambda _ { x } = \frac { v _ { x 0 } } { v _ { x } } \text { and } \lambda _ { y } = \frac { v _ { y 0 } } { v _ { y } }$.
    Consequently, the constant-energy contour can be written in the standard elliptical form
    \begin{equation}
	\frac { [ k _ { a }^\varepsilon - (E-E_0) l _ { a } ] ^ { 2 } } { V _ { a } ^ { 2 } } + \frac { [ k _ { b }^\varepsilon - (E-E_0) l _ { b } ] ^ { 2 } } { V _ { b } ^ { 2 } } = 1\label{ellipse1},\nonumber
\end{equation}
where $V_a$, $V_b$, $l_a$, and $l_b$ are determined by $v_{x0}$, $v_{y0}$, $v_x$, $v_y$, $m$, and $E_0$.
Therefore, the area of the ellipse of a given energy $E$ is $S _ { a b } ( E ) = \pi V _ { a } V _ { b }$. By considering the Jacobian factors of the transformations, the area in the original momentum coordinates is
\begin{equation}
	S ( E ) = \pi R \frac {R ^ { 2 } (E-E_0)^2 - m^2} { v _ { x } v _ { y }\ |\mathrm{det} (1-2\bm{\varepsilon})|},\nonumber
\end{equation}
where $R=(1 - \lambda _ { x } ^ { 2 } - \lambda _ { y } ^ { 2 })^{-1/2}$.
Therefore, the DOS of the general 2D Dirac cone is
\begin{equation}
	\rho ( E ) = \left| \frac { d S ( E ) } { d E } \right| = \left| \frac { 2 \pi (E-E_0) } { v _ { x } v _ { y } \ \mathrm{det} (1-2\bm{\varepsilon})} \right| R^3.\nonumber
\end{equation}
Finally, imposing $|E-E_0|\geq |m/R|$ and using the approximation $\mathrm{det} (1-2\bm{\varepsilon})\approx1$, we obtain the simplified DOS for the strained spin-up Dirac cone:
\begin{equation}
\rho(E)=
\begin{cases}
\left| \alpha E - \beta(\theta)\varepsilon \right|, & |E-\frac{\beta(\theta)\varepsilon}{\alpha}|\geq \frac{1}{2}g|\varepsilon|\\[6pt]
0, & |E-\frac{\beta(\theta)\varepsilon}{\alpha}|< \frac{1}{2}g|\varepsilon|
\end{cases},\nonumber
\end{equation}
where $\alpha = \frac{2\pi R^3}{v_xv_y},\ \beta(\theta) = \alpha\frac{E_0}{\varepsilon},\ g(\theta) = 2\left|\frac{m}{R\varepsilon}\right|$ because $E_0$ and $m$ are linear in signed strain amplitude $\varepsilon$ for small strain with a fixed strain direction $\theta$.
Similarly, the DOS for the strained spin-down Dirac cone can be derived by the transformations $k_i \rightarrow k^{\prime}_i = \sum_{j = x, y} R_{ij}(C)k_j$ and  $\varepsilon_{ij} \rightarrow \varepsilon^{\prime}_{ij} = \sum_{i',j' = x, y} R_{ii'}(C) R_{jj'}(C)\varepsilon_{i'j'}$, where $R_{ij}(C)$ is the element of the representation matrix $\mathbf{R}(C)$ of $C$ \cite{fang2014}.

\subsection*{Strain-induced Lifshitz transition}
We further demonstrate the strain-induced Lifshitz transition through the nonanalyticity of the thermodynamic potential.
First, the low-energy DOS of the Dirac cone $D_\sigma$ can be written as
\begin{equation}
	\rho^\sigma(E)= \alpha\left|E-E_0^\sigma\right|.\nonumber
\end{equation}
Since we focus on the nonanalytic contribution associated with the Van Hove singularity at the Dirac point, analytic background contributions are omitted below.
Then, the corresponding integrated DOS,
\begin{equation}
    \mathcal{N}^\sigma(E) = \int_{-\infty}^{E}{\rho^\sigma(E')\ \mathrm{d}E'},\nonumber
\end{equation}
is
\begin{equation}
    \mathcal{N}^\sigma(E) =\mathrm{sgn}\left(E-E_0^\sigma\right)\frac{1}{2} \alpha \left(E-E_0^\sigma\right)^2.\nonumber
\end{equation}
At zero temperature, the thermodynamic potential, 
\begin{equation}
    \Omega^\sigma(\mu)
= -\int_{-\infty}^{\mu} \mathcal{N}^\sigma(E)\ \mathrm{d}E,\nonumber
\end{equation}
is
\begin{equation}
\Omega^\sigma(\mu)
= -\mathrm{sgn}\left(\mu-E_0^\sigma\right)\frac{1}{6} \alpha \left(\mu-E_0^\sigma\right)^3.\nonumber
\end{equation}
Because $E_0^\uparrow=+\frac{1}{2} s \varepsilon$ and $E_0^\downarrow=-\frac{1}{2}s \varepsilon$, the nonanalytic part of the total thermodynamic potential ($\Omega=\Omega^\uparrow+\Omega^\downarrow$) is
\begin{equation}
\label{eq_ThermPoten}
\Omega(\mu,\varepsilon)=
\begin{cases}
-\dfrac{1}{12}\alpha\,|\mu|\left(4\mu^2+3s^2\varepsilon^2\right),
& |\varepsilon|<\dfrac{2|\mu|}{|s|}, \\[6pt]
-\dfrac{1}{24}\alpha |s|\,|\varepsilon|
\left(12\mu^2+s^2\varepsilon^2\right),
& |\varepsilon|\ge\dfrac{2|\mu|}{|s|}.
\end{cases}\nonumber
\end{equation}
At $\mu=0$, this equation reduces to 
\begin{equation}
    \Omega(0,\varepsilon)=-\dfrac{\alpha |s|^3}{24} |\varepsilon|^3,\nonumber
\end{equation}
which is nonanalytic at $\varepsilon=0$, with a discontinuity in its third derivative with respect to strain. Meanwhile, the Fermi contour changes from point-like at $\varepsilon=0$ to loop-like for any finite $\varepsilon$. Therefore, $\varepsilon=0$ marks a strain-induced Lifshitz transition.

\section*{Funding}

This work is supported by the National Key R$\&$D Program of China (2021YFA1401500) and the Hong Kong Research Grants Council (CRS$\_$HKUST603/25, C6046-24G, 16306722, 16304523 and 16311125).

\section*{Data availability}

The data supporting the findings of this study are available from the corresponding author upon reasonable request.

\section*{Author contributions}
J.L. conceived the idea and supervised the whole project.
X.F., Z.H., F.Y., and J.L. performed the theoretical analysis.
Z.H., M.H., and X.C. performed the first-principles calculations. X.F., F.Y., Z.H., and J.L. wrote the manuscript with input from all the authors.

\section*{Competing interests}

The authors declare no competing interests.

	\bibliography{References}

\end{document}